\documentclass{iaus}
\usepackage{graphicx}

\title[] %% give here short title %%
{Fast rotating stars resulting from binary evolution will often appear to be single}

\author[]   %% give here short author list %%
{S. E. de Mink \and N. Langer \and R. G. Izzard
}

\affiliation{
Argelander-Institut f\"ur Astronomie der Universit\"at Bonn
Germany  \\email: {\tt S.E.deMink@gmail.com}}

\pubyear{2010}
\volume{272}  %% insert here IAU Symposium No.
\pagerange{1--2}
\jname{Active OB stars: structure, evolution, mass loss and critical limit}
\editors{C. Neiner et al.}
\begin{document}

\maketitle

\begin{abstract}
Rapidly rotating stars are readily produced in binary systems. An accreting star in a binary system can be spun up by mass accretion and quickly approach the break-up limit. Mergers between two stars in a binary are expected to result in massive, fast rotating stars.  These rapid rotators may appear as Be or Oe stars or at low metallicity they may be progenitors of long gamma-ray bursts.

Given the high frequency of massive stars in close binaries it seems likely that a large fraction of rapidly rotating stars result from binary interaction.  It is not straightforward to distinguish a a fast rotator that was born as a rapidly rotating single star from a fast rotator that resulted from some kind of binary interaction. Rapidly rotating stars resulting from binary interaction will often appear to be single because the companion tends to be a low mass, low luminosity star in a wide orbit. Alternatively, they became single stars after a merger or disruption of the binary system during the supernova explosion of the primary.  

The absence of evidence for a companion does not guarantee that the system did not experience binary interaction in the past.  If binary interaction is one of the main causes of high stellar rotation rates, the binary fraction is expected to be smaller among fast rotators. How this prediction depend on uncertainties in the physics of the binary interactions  requires further investigation.

\keywords{Stellar evolution, rotation, interacting binaries, critical limit, Be and Oe stars, progenitors of long GRB's}
\end{abstract}

\begin{figure}[t]
% \vspace*{-2.0 cm}
\begin{center}
 \includegraphics[width=\textwidth]{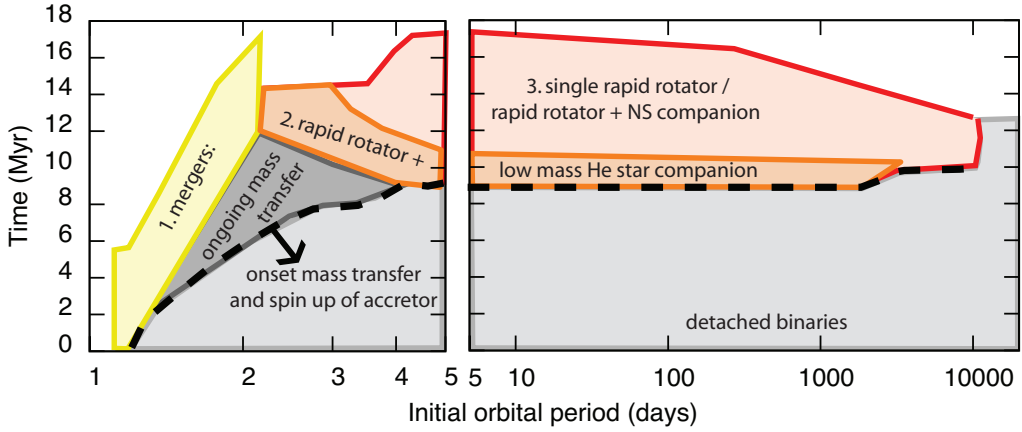} 
% \vspace*{-1.0 cm}
 \caption{Schematic depiction of the evolutionary stages of a 20+15$M_{\odot}$ binary as a function of the initial orbital period ($x$-axis) and age of the system ($y$-axis).  The thick dashed line indicates the onset of mass transfer, which occurs later for wide binaries or is avoided in the widest systems. This diagram indicates that rapid rotators resulting from binary interaction are often single or appear to be single. (1) Mergers in close binaries result in massive, rapidly  rotating, main-sequence stars that are single. (2) Mass transfer drives the accreting star to critical rotation. Initially the companion will be a low mass, low luminosity helium star which will be hard to detect. (3) If the helium star is massive enough it will explode and is likely to disrupt the binary system, leaving its companion behind as a fast spinning single star. This diagram is based on a grid of models computed with an upgraded version of the binary population synthesis code described by Izzard et al (2006) based on Hurley et al (2002),  De Mink et al (in prep.). }
   \label{fig:SdM}
\end{center}
\end{figure}
%\section{Introduction}
%

Rotation gives rise to various phenomena in stars, for example the deformation of fast-rotators to an oblate shape (e.g. Zhao, these proceedings) and mixing processes in stellar interiors (e.g. Zahn,  Ekstr|"om and Brott, these proceedings).  Rotation also seems to be an essential ingredient in the production of disks around classical Be and Oe stars and it may lead to the formation of long GRB progenitors at lower metallicity (Martayan, these proceedings). 

The close binary fraction among massive stars is high. About 45\% of the massive stars in OB associations show radial velocity variations in their spectra (e.g. Sana et al, these proceedings) indicating the presence of a nearby companion.  Binary interaction can lead to rapid rotation rates in various ways, see Fig.~\ref{fig:SdM}. 
%~\\
Given the relative ease with which binary systems produce fast rotating stars and the high fraction of massive stars that are found in binary systems, one may consider the following question: {\it``What fraction of rapid rotators result from binary interaction?''}.  
This question is relevant for various other open questions.  Are Be stars formed in binaries (e.g. Pols et al. 1994, Ekstr\"om et al. 2008)? Are the surface abundances in fast rotating stars signatures of rotational mixing or from binary mass transfer (e.g. Hunter et al. 2008)? Do binaries produce the progenitors of long gamma-ray bursts (Cantiello et al. 2007)?
 
%~\\
Identifying the role of binary interaction in the production of rapid rotators is not straightforward.  The low-luminosity companion star in a post-interaction binary is typically very hard to detect, see Fig~\ref{fig:SdM}. {\bf (I)} Mass transfer will widen the orbit and reduce the mass and luminosity of the originally most massive star.  {\bf (II)} When the companion explodes, the binary system is likely to be disrupted.  The accretor is left behind as a rapidly rotating single star. {\bf (III)} Finally, mergers also produce rapidly rotating single stars.

%~\\
We come to the somewhat counter-intuitive conclusion: fast rotating stars that were produced during binary interactions will, in many or even most cases, appear to be single stars.   In fact, if binary interaction is one of the main causes of high stellar rotation rates, a smaller binary fraction for fast rotators is expected. How this prediction depends on uncertainties in the modeling of the mass transfer phase requires further investigation.

\end{document}